\documentclass[aps,pra,twocolumn,superscriptaddress,amsmath,showpacs,amssymb]{revtex4} 

\usepackage{graphicx}
\usepackage{amsthm,amsmath}

\newcommand{\be}{\begin{eqnarray}}
\newcommand{\ee}{\end{eqnarray}}

\newcommand{\nn}{\nonumber}

\newcommand{\bo}{\boldsymbol}
\newcommand{\lb}{\label}

\newcounter{ichi}
\newcounter{ni}
\newcounter{san}
\newcounter{yon}
\newcounter{go}
\newcounter{roku}
\newcounter{nana}
\newcounter{hati}
\newcounter{kyu}
\begin{document}

\preprint{APS/123-QED}

\title{Tunneling Time of Bose-Einstein Condensates on Real Time Stochastic
 Approach}

\author{K.~Kobayashi}
\email{keita-x@fuji.waseda.jp}
\affiliation{Research Institute for Science and Engineering,
Waseda University, Tokyo 169-8555, Japan}
\author{M.~Inoue}
\affiliation{Department of Electronic and Photonic Systems,
Waseda University, Tokyo 169-8555, Japan}
\author{Y.~Nakamura}
\affiliation{Department of Electronic and Photonic Systems,
Waseda University, Tokyo 169-8555, Japan}
\author{K.~Honda}
\affiliation{Department of Nano-Science and Nano-Engineering, Waseda University, Tokyo 169-8555, Japan}
\author{Y.~Yamanaka}
\affiliation{Department of Electronic and Photonic Systems,
Waseda University, Tokyo 169-8555, Japan}

\date{\today}

\begin{abstract}   
We study tunneling processes of Bose-Einstein condensate (BEC) on the 
real time stochastic approach and reveal some properties of their tunneling time.
An important result is that the tunneling time decreases as 
the repulsive interatomic interaction becomes stronger.
 Furthermore,
 the tunneling time in a strong interaction region is not much affected
 by the potential height and  is represented by an almost constant function.
 We also obtain the other related times such as the hesitating and interaction ones
 and investigate their dependence on the interaction strength.
 Finally, we calculate the mean arrival time of BEC wave packet
 and show the large displacement of its peak position.
\end{abstract}

\pacs{03.75.Lm,42.25.Bs,03.65.Xp,03.65.Sq}
\maketitle

\section{\label{sec:level1}Introduction}
The Bose-Einstein condensates (BECs) of trapped atoms, first realized in
1995 \cite{Cornell,Cornell2,Ketterle2}, are ideal systems for studying
macroscopic quantum phenomena.
 This is because the systems are dilute,
weakly interacting ones and we can easily control the configuration of 
the trap and even the strength of interatomic interaction.
The BEC is described by a macroscopic wavefunction
 with nonlinear effects due to interatomic interactions and
 show the macroscopic tunneling phenomena such as Josephson oscillations.
 Many studies on BEC tunneling through various kinds of potentials \cite{double well1,double well2,tunneling} have already been done. We expect that 
 the BECs give us new insights for nonlinear dynamics and macroscopic tunneling phenomena.

The problem of tunneling time for quantum particle is controversial,
mainly because the time is not represented by an operator but is a parameter
in quantum mechanics.  It has
 been investigated for decades and many definitions of the tunneling time have been proposed 
\cite{phase time,clock,clock2,clock3,Time potential,Time potential2,Time potential3,Bohmian,Bohmian2,path integral,path integral2,Imafuku1}.
 For example, the phase time \cite{phase time} is expressed by
 an energy derivative of phase shift of transmitted wave
from a potential barrier.
 The Larmor time \cite{clock,clock2,clock3} is obtained from the Larmor precession angle caused by
 a magnetic field in the potential barrier region.
The B\"uttiker-Landauer time \cite{Time potential,Time potential2,Time potential3} is defined from transmission coefficient through a barrier with a small oscillation of height.
 There are methods based on the concept of ``particle path" such as
 Bohmian mechanics \cite{Bohmian,Bohmian2}, Feynman path integral \cite{path integral,path integral2}, Nelson's stochastic approach \cite{Imafuku1,Imafuku2,Hara1,Hara2}
 and so on.
 The cold atomic gas systems are promising to 
investigate the tunneling time of quantum particle, since their dynamical behaviors can be
 observed in detail and one can even control several parameters such as
 the configuration of trap,
 the strength of interatomic interaction and
 the internal degree of freedom such as spin.
 
In this paper, we focus on the tunneling time associated with BEC wave function
 and its dependence on the interaction strength.
 But most of the existing definitions of the tunneling time is basically based on the Schr\"odinger equation for one-particle tunneling phenomena.
 The BEC dynamics is well described, particularly near the zero temperature, by the Gross-Pitaevskii (GP) equation, which is the non-linear Schr\"odinger equation. We can not adopt the previous definitions of the tunneling time to the BEC wave
 in a straightforward way.
 Zhang and his co-worker estimated the tunneling time of the BEC wave packet
 qualitatively form its peak motion \cite{GPtime}.
 Their results show that the tunneling time strongly depends on the strength of 
the non-linear interaction.  But the appearance of the negative tunneling time is inevitable
 in their approach.

 For our purpose, we use the Nelson's stochastic approach.
 In Nelson's stochastic quantum mechanics \cite{Nelson1,Nelson2}, 
the random variable $\bo{x}(t)$ representing the motion of a quantum particle
 is subject to  the stochastic differential equations.
 It gives the real time trajectories of the quantum particle as
 sample paths.
 From ensemble of the sample paths, we can reproduce
 the predictions given by the ordinary quantum mechanics.
 Since the Nelson's stochastic approach utilizes the real time trajectories,
 we acquire direct information about the tunneling time.
 We note that it can be extended
 to the system described by the non-linear Schr\"odinger equation \cite{nonlinear1,nonlinear2,Loffrendo}.


In this paper, we explicitly calculate the times related to tunneling phenomena 
 of the BEC wave packets by means of the Nelson's approach.
Their dependence on the interaction strength is an interesting subject.
We note that the GP equation can also be derived by applying the mean field approach directly
within the formulation of the Nelson's quantum mechanics, as will be shown in this paper.
We perform numerical calculations for a quasi one-dimensional system with a rectangular
 potential barrier, that is, one with a barrier in $x$-direction 
and a strong confining harmonic potential in $yz$-plane.
It will be seen from the accumulated sample paths
how the tunneling time of BEC depends on the non-linear interaction strength: it 
decreases as the interaction becomes stronger. Furthermore, it becomes
 constant in a strong interaction region.
 To explain those results, we analyze a simple model with a double well potential. 
 Next, we focus on the hesitating behaviors found in the sample paths,
 which are due to a strong interference between
 the incident and reflecting waves.
 We evaluate the hesitating time and find that
it is much affected by the interaction strength.
 From the calculations of the tunneling and hesitating times,
 we confirm that the Hartman effect  \cite{phase time} is violated
 when the non-linear interaction is switched on.
 Finally, we also calculate the arrival time distribution of the BEC wave packet
and  the mean arrival time from it.
 The result suggests a large acceleration in the motion of the BEC wave packet
 in the presence of a potential barrier and the non-linear interaction.
 Nelson's approach can reproduce the arrival time distribution.
 We argue that the displacement of the peak position 
accounts for the  ``acceleration".

This paper is organized as follows.
 In Sec.~II, we review the Nelson's quantum mechanics and 
 derive the GP equation directly in the stochastic approach, using the mean field approach.
 In Sec.~III, we numerically calculate the tunneling time in the stochastic approach.
 In Sec.~IV, we discuss the mean arrival time of the BEC wave packet.
 Section V is devoted to summary.

\section{\label{sec:level2}Real Time Stochastic Process and Gross-Pitaevskii Equation}
In this section, we briefly review the Nelson's stochastic quantum mechanics
 with $N$ degrees of freedom \cite{Nelson1,Nelson2,Loffrendo} and show its equivalence to the Schr\"odinger equation.
 Next, applying the mean field approach to the dynamical and kinematical equations,
 we derive the GP equation.

The first assumption of the Nelson's stochastic quantization is to set up the 
stochastic differential equations for the particle position $\bo{x}_{i}(t)$
for $i=1,\cdots,N$ as 
\be
&&d\bo{x}_{i}(t)=
\bo{b}(\bo{X}(t),t)dt+\sqrt{\frac{\hbar}{m_{i}}}d\bo{W}_{i}\lb{eq:FSE} \,,\\
&&d\bo{x}_{i}(t)
=\bo{x}_{i}(t+dt)-\bo{x}_{i}(t)\,, \nn
\ee
for forward time evolution and 
\be
&&d\bo{x}_{i}(t)=\bo{b}_{*}(\bo{X}(t),t)dt+\sqrt{\frac{\hbar}{m_{i}}}d\bo{W}_{i*} \lb{eq:BSE}\,,\\
&&d\bo{x}_{i}(t)=\bo{x}_{i}(t)-\bo{x}_{i}(t-dt)\,,\nn
\ee
for backward one with $dt>0$, particle masses $m_{i}$ and
 the notation, $\bo{X}(t)\equiv\left(\bo{x}_{1}(t),\cdots,\bo{x}_{N}(t)\right)$\,.
 Here $\bo{W}_{i}$ and $\bo{W}_{i*}$ are three-dimensional independent
standard Wiener processes
\be
&&E[dW_{l,i}(t)]=E[dW_{l,i*}(t)]=0\,, \\
&& E[dW_{l_{1},i}(t)dW_{l_{2},j}(t)]=E[dW_{*l_{1},i}(t)dW_{*l_{2},j}(t)]\nn \\
&&=\delta_{l_{1}l_{2}}\delta_{ij}dt\,,
\ee 
where $E[\cdots]$ means an ensemble average and indices $l$ and $i$ represent
 $l=x,y,z$ and $i=1,\cdots,N$\,, respectively.

The second assumption is the Newton-Nelson's equation of motion as follows:
\be
m_{i}\bo{a}_{i}(t)=-\nabla_{i}V(\bo{X}(t),t) \lb{eq:Newton}\,,
\ee
where $V(\bo{X}(t),t)$ represents the sum of the external potential $V_{\rm ex}(\bo{x}_{i},t)$
 and interaction potential $V_{\rm int}(\bo{x}_{i}-\bo{x}_{j})$ as $V=\sum_{i}V_{\rm ex}(\bo{x}_{i},t)+(1/2)\sum_{i\neq j}V_{\rm int}(\bo{x}_{i}-\bo{x}_{j})$\,. The mean acceleration
 $\bo{a}_{i}(t)$ is defined as
\be
\bo{a}_{i}(t)=\frac{1}{2}(DD_{*}+D_{*}D)\bo{x}_{i}(t)\,,
\ee
with the mean forward time derivative $D$\,,
\be
Df(t)=\lim_{dt\to 0+}E\left[\frac{f(\bo{X}(t+dt))
-f(\bo{X}(t))}{dt}\,\bigg{|}\,\bo{X}(t)\right]\,,
\ee
and the mean backward one $D_{*}$\,, 
\be
D_{*}f(t)=\lim_{dt\to 0+}E\left[\frac{f(\bo{X}(t))-f(\bo{X}(t-dt))}{dt}
\,\bigg{|}\,\bo{X}(t)\right]\,,
\ee
 where $E[\cdots|\bo{X}(t)]$ means the conditional expectation.
Let us define the current and osmotic velocities as
$
\bo{v}=(\bo{b}+\bo{b}_{*})/2
$ and
$
\bo{u}=(\bo{b}-\bo{b}_{*})/2
$\,. 
Then, the Newton-Nelson's equation of motion (\ref{eq:Newton})
 becomes the equation for the current velocity:
\be
\frac{\partial}{\partial t}\bo{v}_{i} 
&=&\sum_{j}\left[\frac{\hbar}{2m_{j}}\nabla_{j}^{2}\bo{u}_{i}
+(\bo{u}_{j}\cdot \nabla_{j})\bo{u}_{i}-(\bo{v}_{j}\cdot \nabla_{j})\bo{v}_{i}
\right]\nn\\
&&\quad-\frac{1}{m}\nabla_{i} V \,.
\lb{eq:NelsonCeq}
\ee

The stochastic processes of the random variables $\bo{x}_i(t)$ in Eqs.(\ref{eq:FSE})
 and (\ref{eq:BSE}) are equivalently formulated by means of the distribution
function $P(\bo{X},t)$ which satisfies the Fokker-Planck equations,
\be
\frac{\partial}{\partial t}P=-\sum_{i}\nabla_{i}\cdot(\bo{b}_{i}P)+\sum_{i}\frac{\hbar}{2m_{i}}\nabla_{i}^{2}P\,,
 \lb{eq:fFP}
\ee
for forward time and
\be
\frac{\partial}{\partial t}P=-\sum_{i}\nabla_{i}\cdot(\bo{b}_{i*}P)-\sum_{i}\frac{\hbar}{2m_{i}}\nabla_{i}^{2}P\,,
 \lb{eq:bFP}
\ee
for backward one. 
 It is known that the forward and backward velocities have a certain relation
$
\bo{b}_{i}-\bo{b}_{*i}=\frac{\hbar}{m_{i}}\nabla_{i}\ln P
$ \cite{Nelson2}.
From Eqs.(\ref{eq:fFP}), (\ref{eq:bFP}) and $\bo{u}_{i}=\frac{\bo{b}_{i}-\bo{b}_{*i}}{2}=\frac{\hbar}{2m_{i}}\nabla_{i}\ln P$\,, one derives
 the equation for the osmotic velocity as
\be
\frac{\partial}{\partial t}\bo{u}_{i}
=-\sum_{j}\left[\frac{\hbar}{2m_{j}}\nabla_{j}^{2}\bo{v}_{i}-\nabla_{i}(\bo{u}_{j}\cdot\bo{v}_{j})\right]\,. 
\lb{eq:NelsonOeq}
\ee
Equations (\ref{eq:NelsonCeq}) and (\ref{eq:NelsonOeq})
 are called the dynamical and kinematical equations, respectively, 
 and their combination can be transformed into the Schr\"odinger equation,
\be
i\hbar\frac{\partial}{\partial t}\Psi(\bo{X},t)
=\left(-\sum_{i}\frac{\hbar^{2}}{2m_{i}}\nabla_{i}^{2}+V(\bo{X},t)\right)\Psi(\bo{X},t)\,, \lb{eq:Scheq}
\ee
with the substitutions, 
$
\bo{v}_{i}=\frac{\hbar}{m_{i}}{\rm Im}[\nabla_{i}\ln\Psi]
$
 and 
$
\bo{u}_{i}=\frac{\hbar}{m_{i}}{\rm Re}[\nabla_{i}\ln\Psi]
$\,.
 The probability of the particle position $P(\bo{X},t)$ corresponds to the 
 absolute square of the Schr\"odinger amplitude $P(\bo{X},t)=|\Psi(\bo{X},t)|^{2}$\,.
 Thus, the stochastic differential equations (\ref{eq:FSE}) and (\ref{eq:BSE}) are equivalent
 to the Schr\"odinger equation. 

Next, we apply the mean field approach to the Nelson's stochastic quantum mechanics
 and derive the GP equation.
 We average the $i$-th dynamical and kinematical equations
 over the variables $(\bo{x}_{1},\cdots,\bo{x}_{i-1},\bo{x}_{i+1},\cdots,\bo{x}_{N})$
 as
\be
&&\frac{\partial}{\partial t}E_{\bo{x}_{i}}\left[\bo{v}_{i}\right] \nn \\
&=&\sum_{j}E_{\bo{x}_{i}}\left[
\frac{\hbar}{2m}\nabla_{j}^{2}\bo{u}_{i}
+(\bo{u}_{j}\cdot \nabla_{j})\bo{u}_{i}-(\bo{v}_{j}\cdot \nabla_{j})\bo{v}_{i}
\right]\nn\\
&&\quad-E_{\bo{x}_{i}}\left[\frac{1}{m}\nabla_{i} V\right] \,,
\lb{eq:meanNelsonCeq} \\
&&\frac{\partial}{\partial t}E_{\bo{x}_{i}}\left[\bo{u}_{i}\right] \nn \\
&=&-\sum_{j}E_{\bo{x}_{i}}\left[\frac{\hbar}{2m}\nabla_{j}^{2}\bo{v}_{i}-\nabla_{i}(\bo{u}_{j}\cdot\bo{v}_{j})\right]\,, 
\lb{eq:meanNelsonOeq}
\ee
where $E_{\bo{x}_{i}}[\cdots]$ means the conditional expectation as
\be
E_{\bo{x}_{i}}[F(\bo{X})]&=&\int \prod_{j\neq i}d\bo{x}_{j} F(\bo{X})\rho_{i}(\bo{X},t)\,, \\
\rho_{i}(\bo{X},t)&=&\frac{P(\bo{X},t)}{\int \prod_{j\neq i}d\bo{x}_{j}P(\bo{X},t)}\,.
\ee
Here, we apply the mean field ansatz and put the factorized probability distribution for
 the particles position as
 $P(\bo{X},t)=\prod_{j} P_{j}(\bo{x}_{j},t)$\,.
Since a BEC phase is under consideration,
every particle should have
the same probability distribution $P_{1}(\bo{x},t)=\cdots=P_{N}(\bo{x},t)={\bar P}(\bo{x},t)$\,.
 Then, all the current and osmotic velocities should be in the same forms irrespective of the index $i$, $\bo{v}_{i}(\bo{X},t)=\bo{\bar{v}}(\bo{x}_{i},t)\equiv \bo{\bar{v}}_i$ and
 $\bo{u}_{i}(\bo{X},t)=\bo{\bar{u}}(\bo{x}_{i},t)\equiv \bo{\bar{u}}_{i}$\,, and Eqs.(\ref{eq:meanNelsonCeq}) and (\ref{eq:meanNelsonOeq}) become
\be
\frac{\partial}{\partial t}\bo{\bar{v}}_{i}&=&
\frac{\hbar}{2m}\nabla_{i}^{2}\bo{\bar{u}}_{i}
+(\bo{u}_{i}\cdot \nabla_{i})\bo{\bar{u}}_{i}-(\bo{\bar{v}}_{i}\cdot \nabla_{i})\bo{\bar{v}}_{i} \nn \\
&&-\frac{1}{m}\int \prod_{j\neq i}d\bo{x}_{j}\rho_{i}\nabla_{i}V \,,
\lb{eq:GPNelsonCeq} \\
\frac{\partial}{\partial t}\bo{\bar{u}}_{i}
&=&-\frac{\hbar}{2m}\nabla_{i}^{2}\bo{\bar{v}}_{i}+\nabla_{i}(\bo{\bar{u}}_{i}\cdot\bo{\bar{v}}_{i})\,. 
\lb{eq:GPNelsonOeq}
\ee
Now, we take  the contact-type interaction potential,
\be
V_{\rm int}(\bo{x}_{i}-\bo{x}_{j})
=g\delta(\bo{x}_{i}-\bo{x}_{j})
\ee
where $g=4\pi\hbar^{2}a_{\rm s}/m$
 with $s$-wave scattering length $a_{\rm s}$\,. Note that a repulsive interaction $g>0$ is 
considered for BEC.
Utilizing the transformations 
$\bo{\bar{v}}_{i}=\frac{\hbar}{m}{\rm Im}[\nabla_{i}\ln\psi]$ and
$\bo{\bar{u}}_{i}=\frac{\hbar}{m}{\rm Re}[\nabla_{i}\ln\psi]$\,,
we obtain the non-linear Schr\"odinger equation from Eqs. (\ref{eq:GPNelsonCeq}) and (\ref{eq:GPNelsonOeq}) as
\be
i\hbar\frac{\partial }{\partial t}\psi(\bo{x},t)=\left[-\frac{\hbar^{2}}{2m}
\nabla^2+V_{\rm ex}+g(N-1)
|\psi|^{2}
\right]\psi(\bo{x},t)\,.\nn\\
\ee
With $N\simeq N-1$ and $\Psi=\sqrt{N}\psi$, 
it is the GP
 equation. The corresponding stochastic differential equations are given by
\be
&&d\bo{x}(t)=
\left[\bo{\bar{v}}(\bo{x}(t),t)+\bo{\bar{u}}(\bo{x}(t),t)\right]dt+\sqrt{\frac{\hbar}{m}}d\bo{W}\lb{eq:FGSE} \,,\\
&&d\bo{x}(t)=
\left[\bo{\bar{v}}(\bo{x}(t),t)-\bo{\bar{u}}(\bo{x}(t),t)\right]dt+\sqrt{\frac{\hbar}{m}}d\bo{W}_{*} \,.\lb{eq:BGSE}
\ee

\section{Tunneling Time of Bose-Einstein Condensate}

We consider a system with an external potential in $x$-direction, $V_{\rm ex}(x)$\,,
and a strong confining harmonic potential in $yz$-plane.
 Suppose that the transverse confinement is very strong and the transverse excitations
are forbidden.  
 Then the total wave function $\Psi(\bo{x},t)$ can be approximated as
\be
\Psi(\bo{x},t)\simeq\psi(x,t)\phi(r_{\bot})e^{-i\omega_{\bot}t}\,,
\ee
where $\psi(x,t)$ is the wave function in $x$-direction mode and
 $\phi(r_{\bot})$ ($r_{\bot}=\sqrt{y^2+z^2}$) is that of the transverse ground state.
This system is called a quasi one-dimensional system, and we perform 
numerical calculations with a rectangular barrier
 potential for $V_{\rm ex}(x)$\, .

 For the weak interaction, $\phi(r_{\bot})$ can be replaced by the
 Gaussian form
\be
\phi(r_{\bot})=\sqrt{\frac{m\omega_{\bot}}{\pi\hbar}}\exp\left(-\frac{m\omega_{\bot}r_{\bot}^{2}}{2\hbar}\right)\,.
\ee
Then, the GP equation for $\psi(x,t)$ is given by
\be
i\frac{\partial }{\partial t}\psi(x,t)=\left[-\frac{\partial^{2}}{\partial x^{2}}+V_{\rm ex}(x)+g
|\psi(x,t)|^{2}
\right]\psi(x,t)\,,
\ee
with $x$ in a unit of $a_{\bot}=\sqrt{\hbar/m\omega_{\bot}}$, $t$ in a unit of $\omega_{\bot}^{-1}$, 
the dimensionless wave function $\psi/a_{\bot} \rightarrow \psi$\,,
 and the nonlinear interaction $2a_{s}N/a_{\bot} \rightarrow g$\,.
We consider a rectangular barrier potential as 
\be
V_{\rm ex}(x)=\left\{
\begin{array}{ccc}
0\,, &x<0& (\text{{\rm region I}})\,,\\
V_{0}\,, &\qquad0\le x\le d& (\text{{\rm region II}})\,,\\
0\,, &d< x& (\text{{\rm region III}})\,,
\end{array}
\right. 
\ee
with the unit of $\hbar\omega_{\bot}$\,.
We assume that the initial wave packet with a variance $(\Delta x)^{2}$
 has the Gaussian form as
\be
\psi(x,0)&=&\left(\frac{1}{2\pi(\Delta x)^{2}}\right)^{1/4}\exp\left[
-\frac{(x-x_{0})^{2}}{4(\Delta x)^{2}}\right.\nn\\
&&\quad+ik_{0}(x-x_{0})
\Bigg]\,,\nn\\
\ee
where $x_{0}$ and $k_{0}$ are the initial center position
 and momentum of the wave packet, respectively.
\begin{figure}[h]
\begin{center}
\includegraphics[width=1.00\linewidth]{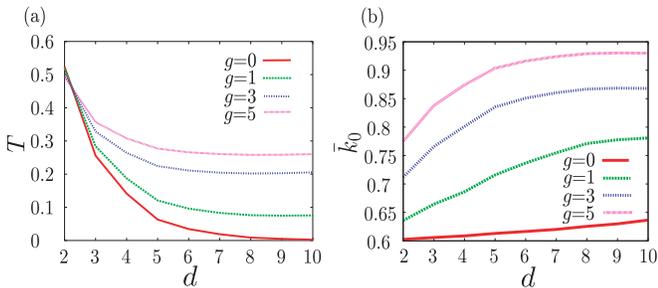}
\end{center}
\caption{(Color Online) (a) The transmission rate of
 the BEC wave packet versus the potential width for the interaction strength $g=0,1,3,5$\,.
 (b) The center momentum of the transmitted wave packet versus the potential width
 for $g=0,1,3,5$\,.
 The other parameters are $V_{0}=1.4(k_{0}^{2}/2)$ and $k_{0}=0.6$\,.
}
\label{trans}
\end{figure}

\begin{figure}[h]
\begin{center}
\includegraphics[width=0.50\linewidth]{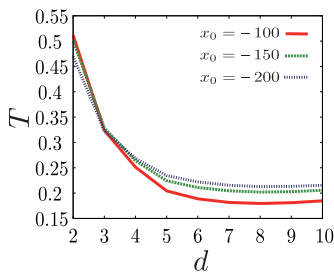}
\end{center}
\caption{(Color Online) The transmission rate of
 the BEC wave packet versus the potential width for the initial position
 $x_{0}=-100, -150, -200$\,.
The other parameters are $g=3$\,, $V_{0}=1.4(k_{0}^{2}/2)$ and $k_{0}=0.6$\,.
}
\label{trans2}
\end{figure}

We now investigate the tunneling of a BEC wave packet.
 In Fig.\ref{trans}-(a), we calculate its transmission rate as
\be
T=\int_{d}^{\infty}\! dx \, |\psi_{\rm T}(x,t_{\rm f})|^{2}\,,
\ee
where $\psi_{\rm T}$ is the transmitted wave packet
 and $t_{\rm f}$ is the final time of the scattering problem.
 We can find that
 the transmission rates depend very much 
on the nonlinear interaction strength $g$\,.
 This dependence is mainly because the repulsive interaction converts
part of the interaction energy into the kinetic one 
particularly in high-density regions.
The energy conversion is also responsible for the dependence of
 the center momentum
\be
\bar{k}_{0}=\int\! dk \, k|\psi_{T}(k)|^{2}\,,
\ee
on $g$, as is shwon in Fig.\ref{trans}-(b)\,.
 Finally, we refer to the initial position dependence of the transmission rate.
 The interaction energy is also being transformed into the kinetic energy
 even outside of the potential barrier and broadens the width of the wave packet.
 It affects the transmission rate
 in Fig.\ref{trans2} for $x_{0}=-100, -150, -200$\,, 
but the dependence is not significant in our choice of the parameters.
 So we show only the results for the initial position $x_{0}=-150$\, below.

\begin{figure}[h]
\begin{center}
\includegraphics[width=0.80\linewidth]{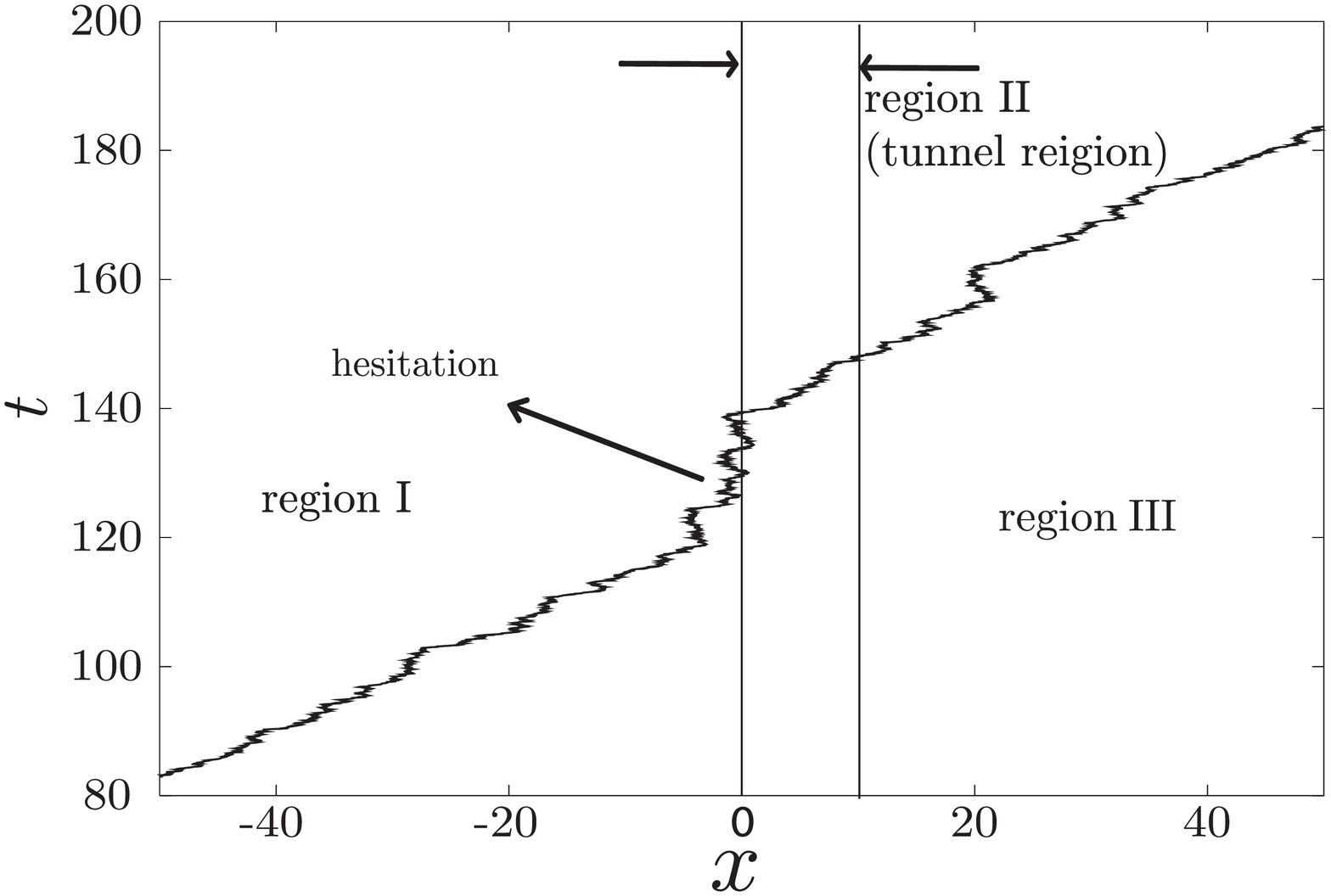}
\end{center}
\caption{\footnotesize{ 
Typical transmission sample path with hesitation.
}}
\label{Samplepath}
\end{figure}
Let us turn to calculate
the tunneling time of the BEC wave packet.
 Figure~\ref{Samplepath} shows a typical transmission sample path,
 calculated by Eq.(\ref{eq:BGSE}) with 
``backward time evolution method" \cite{Imafuku1,Imafuku2}.
 The tunneling time $\tau_{{\rm T}}$ is defined as
the averaged time interval in which the random variables $x(t)$ stay
in the barrier region II, 
\be
\tau_{{\rm T}}&=&\sum_{l}^{N_{\rm Sample}}\frac{1}{N_{\rm Sample}}
\int_{t_{l}=0}^{t_{f}}\Theta(x_{l}(t)) dt\,, \\
\Theta(x)&=&\left\{
\begin{array}{ccc}
0\,, &x<0 \\
1\,, &\quad 0\le x\le d \\
0\,, &d< x \,,
\end{array}
\right. 
\ee
where $t_{i}=0$ and $t_{f}$ are the initial and final times for the scattering problem, respectively.
As shown in Fig.\ref{Samplepath}, the random variable $x(t)$ of the transmission sample path stays in front of the potential barrier \cite{Imafuku1}.
 This hesitating phenomenon is due to a strong interference between the incident and reflecting waves.
We define the hesitating time $\tau_{{\rm H}}$ as the averaged time interval in which the random variables
 $x(t)$ pass through some region in front of the potential barrier $(-d_{H}\le x <0)$\,.
The interaction time $\tau_{{\rm I}}$ is also defined  as a sum of 
the tunneling time $\tau_{{\rm T}}$ and the hesitating one $\tau_{{\rm H}}$\,.
Thus the interaction time $\tau_{{\rm I}}=\tau_{{\rm T}}+\tau_{{\rm H}}$ represents the passage time through the region $(-d_{H}\le x <d)$\,.

\begin{figure}[h]
\begin{center}
\includegraphics[width=1.00\linewidth]{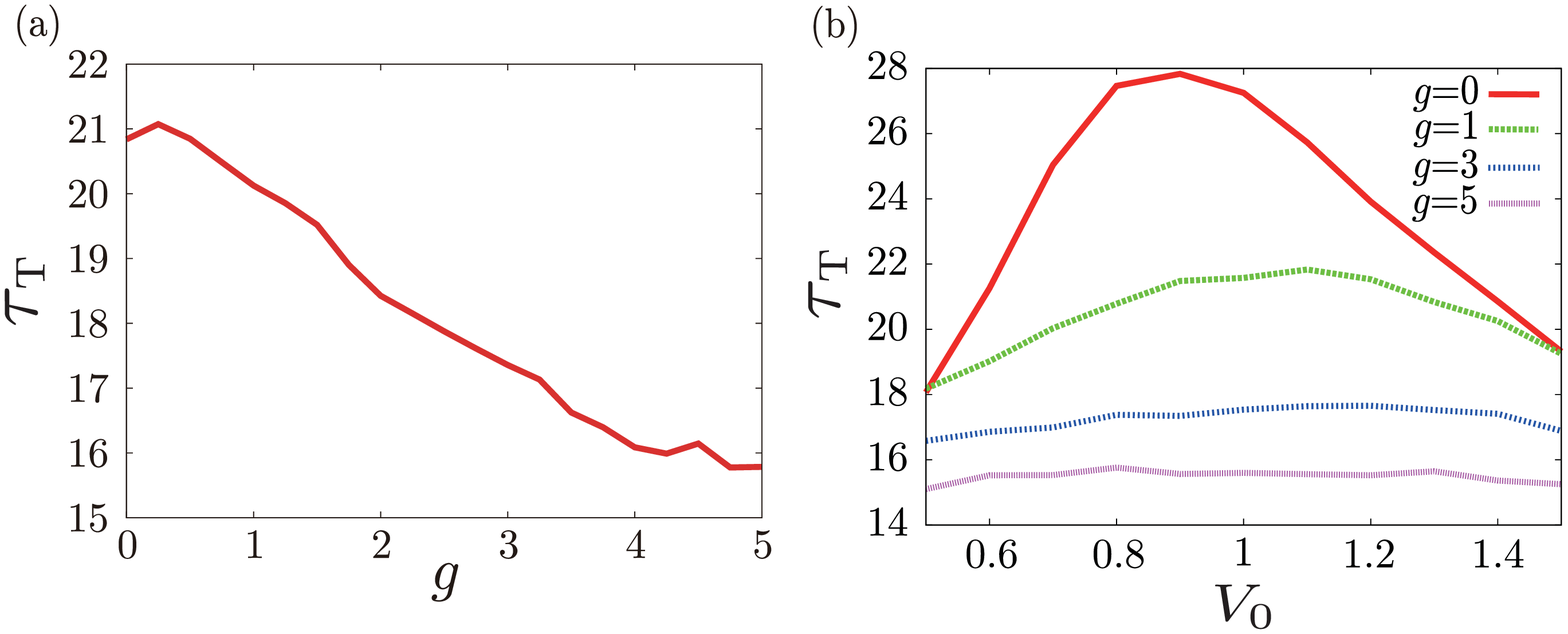}
\end{center}
\caption{(Color Online) (a) The tunneling time of
 the BEC wave packet versus the interaction strength $g$ with the 
parameters $d=7$, $V_{0}=1.4(k_{0}^{2}/2)$ and $k_{0}=0.6$.
(b)  The tunneling time of the
 BEC wave packet versus the potential height $V_{0}/(k_{0}^{2}/2)$ for
 $g=0,1,3,5$\,.
The other parameters are $d=7$ and $k_{0}=0.6$\,.
}
\label{tunneling time1}
\end{figure}

 We study how the interaction strength $g$ affects the tunneling time.
 Figure~\ref{tunneling time1}-(a) shows a the behavior of the tunneling time
 versus $g$.
 We can see that the tunneling time decreases as $g$ increases. 
The nonlinear repulsive interaction accelerates the motion of the BEC wave packet
 in the potential region.
 Figure~\ref{tunneling time1}-(b) shows the behavior of the tunneling time
 versus the potential height.
 The tunneling time with $g=0$ increases first as the potential becomes higher, 
which can be understood intuitively, but  shifts to a decrease for the high potential.
 The latter behavior can be explained by the B\"uttiker-Landauer time
 $\tau=md/(\hbar\sqrt{2(V_{0}-E)})$\,,
 which is also obtained by the Nelson stochastic approach for high and 
 wide potential barrier with $g=0$ \cite{Imafuku1}.
 On the other hand, the tunneling time does not vary much with the potential height
$V_0$ in case of the strong interaction, and approaches to a constant value.
 Our results imply that the tunneling time of the BEC wave packet
 mainly depends on the interaction strength $g$, but not on the potential height.

To explain the $V_0$ independence of the tunnneling time for large $g$,
 we take a simple model of BECs in a double well potential.
 We suppose that  two BECs are initially in a stationary state
 with the condensate particle number difference $\Delta N(0)=0$ and the 
phase difference $\Delta \phi=0$\,.
 Then, we add $\Delta N_{0}$ particles to the left well, $\Delta N(0)=-\Delta N_{0}$, and 
 this number difference induces the tunnel current
 form the left well to the right one.
Then the motion of the BECs
 between the two wells are described by the simultaneous equations \cite{pethick}
\be
&&\frac{d}{dt}\Delta N(t) =2\frac{JN}{\hbar}\Delta\phi(t) \lb{eq;JP1}\,,\\
&&\frac{d}{dt}\Delta \phi(t) =-\left(U+\frac{2J}{N}\right)\Delta N(t) \,,\lb{eq;JP2}
\ee
where $N$, $J$, and $U$ represent the total condensate particle number, 
the tunnel coefficient, and the interaction constant, respectively.
 It is assumed in derivation of Eqs.(\ref{eq;JP1}) and (\ref{eq;JP2})
that $\Delta N_{0}/N$ and $\Delta\phi$
 are small.
 Their solutions  are given by
\be
\Delta N &=&-\Delta N_{0}\cos\left(\omega t\right) \,,\\
\Delta \phi&=& \Delta N_{0}\frac{\hbar\omega}{2JN}\sin\left(\omega t\right)\,,
\ee
with the frequency $\hbar\omega=\sqrt{2J(NU+2J)}$\,.
Since we are not interested in an oscillation of the two BECs here
but only in  the tunneling from the left to the right,
we consider only for $\omega t\ll  1$ and ignore the order $\mathcal{O}(t^{2})$.
Then, we obtain
\begin{align}
	\Delta \phi 
	&\simeq \frac{\Delta N_0}{N} \frac{h\omega^2t}{2J} 
	\simeq 
	\begin{cases}
	\frac{\Delta N_0}{N} \frac{2Jt}{\hbar} & (\text{for\;} U\ll J/N)\,, \\[5pt]
	\frac{\Delta N_0}{N} \frac{NUt}{\hbar} & (\text{for\;} U\gg J/N)\,.
	\end{cases}
\end{align}
The phase difference $\Delta\phi$ corresponds to the tunneling current velocity, 
so larger $\Delta\phi$ implies smaller tunneling time.
One notices the  following two features of the phase difference $\Delta\phi$.
First, it increases monotonically as $U$ does.
Second, while it depends on $J$ in the weak interaction limit $U\ll J/N$, 
it becomes independent in the strong interaction limit $U\gg J/N$.
As $U$ and $J$ in the toy model with the double well
can be identified with $g$ and the potential height
$V_0$ in the model of the potential barrier, respectively,
the arguments just above explain the behaviors of the tunneling time
in Fig.\ref{tunneling time1}\,.

\begin{figure}[h]
\begin{center}
\includegraphics[width=1.00\linewidth]{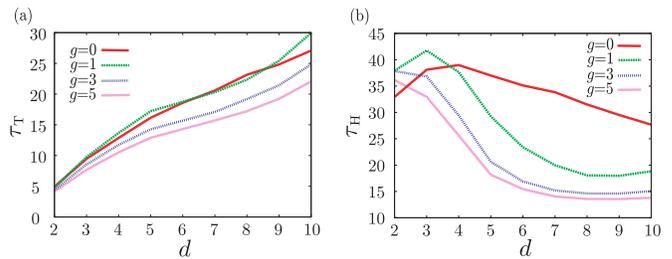}
\end{center}
\caption{(Color Online)(a) The tunneling time of the BEC wave packet versus the potential width $d$ for $g=0,1,3,5$\,. (b) The hesitating time of the BEC wave packet versus
 $d$ for $d_{\rm H}=10$ and $g=0,1,3,5$\,.
The other parameters are $V_{0}=1.4(k_{0}^{2}/2)$ and $k_{0}=0.6$\,.}
\label{tunneling time2}
\end{figure}

\begin{figure}[h]
\begin{center}
\includegraphics[width=0.50\linewidth]{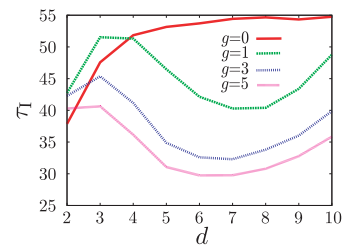}
\end{center}
\caption{(Color Online) The interaction time of the BEC wave packet versus the
potential width $d$ for $g=0,1,3,5$\,. The other parameters 
are $V_{0}=1.4(k_{0}^{2}/2)$ and $k_{0}=0.6$\,.}
\label{tunneling time3}
\end{figure}
Finally, we investigate the properties of the hesitating
 and interaction times
 versus the potential width $d$\,, which are shown in Fig.\ref{tunneling time2}.
 For $g=0$, while the tunneling time increases monotonically,
 the hesitating time decreases for large $d$.
 This is because that the particle of 
the transmission sample path for thick potential
tends to have higher velocity and therefore to 
pass through the region in front of the potential barrier faster
 \cite{AOKI}. 
 As $g$ becomes larger,
 the tunneling and hesitating times decrease.
 One can see that the hesitating time is affected by the interaction
 strength much more than the tunneling time, as in Fig.\ref{tunneling time2}.
 This result can be explained as follows:
 The incident and reflecting waves make 
 a strong interference in front of the potential barrier and
 create the high density region. 
 There the non-linear repulsive interaction term $g|\psi|^{2}$ 
 contributes strongly to the behavior of the sample path.
 Next, we refer to the dependence of
 the interaction time on $d$, as in Fig.\ref{tunneling time3}\,.
 It is predicted, based on the method of phase time \cite{phase time},
 that the ``tunneling time" for thick-enough barrier
 becomes independent of the barrier length for non-interacting systems,
 which is known as the Hartman effect.
 In the Nelson's stochastic approach, the tunneling time with $g=0$ grows
 but the hesitating time decreases, and their sum, the interaction time,
 seems to approach to a constant value, as in Fig.\ref{tunneling time3}.
 This corresponds to the Hartman effect
 in Nelson's stochastic approach.
 We remark that the Hartman effect is apparently violated in the presence of the
 nonlinear interaction, $g\neq 0$\,.
 The violation of the Hartman effect for non-linear interaction has also 
been pointed out in Ref.~\cite{GPtime}\,.

\section{Arrival Time of the BEC Wave Packet}
As seen in Fig.\ref{trans}-(b)\,,
 the center momentum for the transmitted wave packet becomes larger than 
 that for the incident one.
 In tunneling process, the non-linear interaction reduces
the tunneling and hesitating times. 
 The acceleration of the quantum particle
 in the presence of the potential barrier and nonlinear interaction
accounts for these results.
 In this section, we investigate the acceleration of the wave packet
 in view of the mean arrival time \cite{AOKI}.

 Introduce the arrival time distribution at the position $x$ as
\be
P_{x}(t)=\frac{|\psi(x,t)|^{2}}{\int_{t_{i}=0}^{t_{f}}dt|\psi(x,t)|^{2}}\,,
\ee
and the mean arrival time $T_{x}$ as
\be
T_{x}=\int_{t_{i}=0}^{t_{f}}\! dt \, tP_{x}(t)\,.
\ee
One can calculate the difference between the mean arrival times with and without a
potential barrier,
\be
\Delta T_{x}=T_{x}^{\rm tunnel}-T_{x}^{\rm free} \,.
\ee

Due to the nonlinear interaction,
the center of the momentum for the transmitted wave packet becomes large
in Fig.\ref{trans}-(b) 
 and the mean arrival time $T_{x}$ should reflect this effect.
 In order to study the
 acceleration of the BEC wave packet in the potential barrier,
 we focus on the mean arrival time at the potential barrier edge $x=d$\,.

\begin{figure}[h]
\begin{center}
\includegraphics[width=1.0\linewidth]{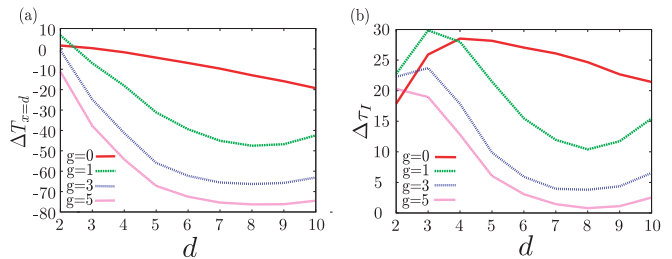}
\end{center}
\caption{(Color Online) (a) The difference between the mean arrival times of the BEC wave packets with and without  a potential barrier
 versus the potential width $d$ for $g=0,1,3,5$\,.
(b) The difference between the interaction times of the BEC wave packets
 with and without a potential barrier versus the potential width $d$ for $g=0,1,3,5$\,.
 The other parameters are $V_{0}=1.4(k_{0}^{2}/2)$ and $k_{0}=0.6$\,.}
\label{arrivaltime}
\end{figure}

\begin{figure}[h]
\begin{center}
\includegraphics[width=1.0\linewidth]{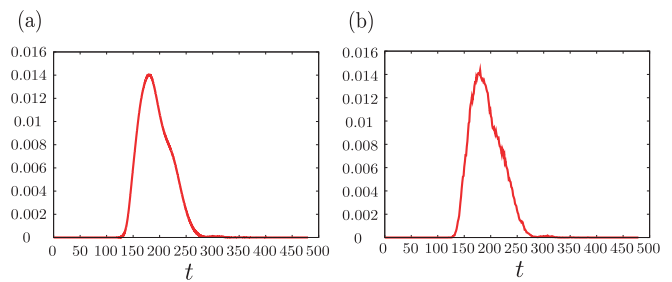}
\end{center}
\caption{(Color Online)(a) The arrival time distribution of the BEC  wave packet $P_{x=10}(t)$.
(b) The histogram obtained from an ensemble of transmitted sample paths
 at $x=10$\,.
 The other parameters are $g=5$, $d=10$, $V_{0}=1.4(k_{0}^{2}/2)$ and $k_{0}=0.6$\,. }
\label{TProb}
\end{figure}

The results of $\Delta T_{x=d}$ are shown in Fig.\ref{arrivaltime}-(a)\,.
 At first, we see that $\Delta T_{x=d}$ can become negative and that
 the non-linear interaction gives rise to a large $|\Delta T_{x=d}|$\,,
which strongly suggests a big acceleration inside the potential barrier region. 
For comparison, we also show the difference between
 the interaction times with a potential barrier $\tau_{\rm I}$ and
 without it $\tau_{\rm I}^{\rm free}$
 in Fig.\ref{arrivaltime}-(b)\,. The difference
 $\Delta\tau_{\rm I}=\tau_{\rm I}-\tau_{\rm I}^{\rm free}$
 becomes small as the interaction strength goes up,
 but does not become negative in contrast to
$\Delta T_{x=d}$.
 It indicates that the velocity of the transmitted wave packet does not
 exceed that of the free wave packet.
 Although  the above results sound paradoxical at first glance,
 the Nelson's method can reproduce
 physical quantities in quantum mechanics and the arrival time
 distribution can
 actually be obtained from  the transmitted sample paths (Fig.\ref{TProb}).

\begin{figure}[h]
\begin{center}
\includegraphics[width=1.0\linewidth]{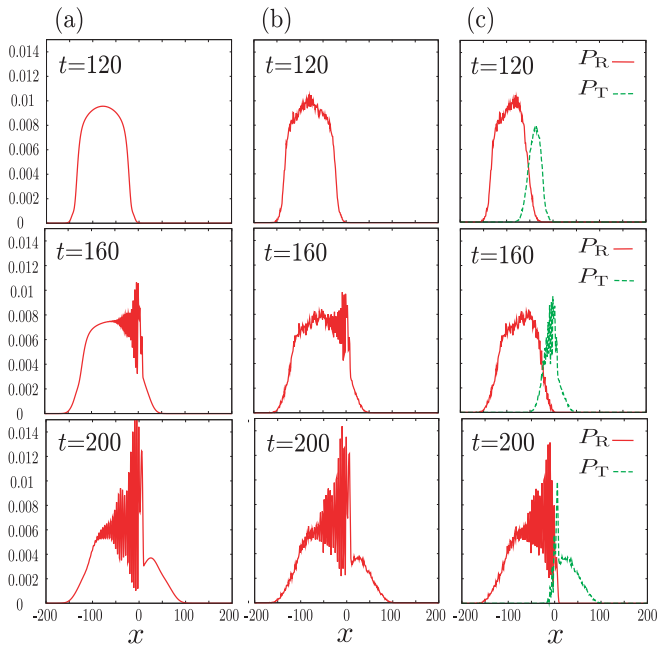}
\end{center}
\caption{(Color Online) (a) The absolute square of the amplitude $\psi(x,t)$
 for $t=120,160,200$\,.
(b) The probability distribution obtained from ensembles of
 the transmitted and reflected sample paths for $t=120,160,200$\,.
(c) The probability distributions of the transmitted and reflected components, 
which are obtained from ensembles of the transmitted and reflected sample paths 
for $t=120,160,200$\,. 
The other parameters are $g=5$, $d=10$, $V_{0}=1.4(k_{0}^{2}/2)$ and $k_{0}=0.6$\,.
}
\label{wavemotion}
\end{figure}

The key to understand these results is a displacement of the peak position
of the wave packet. 
Figure ~\ref{wavemotion}-(a) shows the wave packets at
 $t=120, 160, 200$\,.
 It tells us that 
 before the peak of the incident wave packets reaches the potential barrier
 $x=0$ the peak of the transmitted one appears at $t=200$\,.
 Furthermore, only the front part of the incident wave packet seems to
 contribute to the transmission.
 This situation becomes clear in the Nelson's stochastic interpretation.
 From the transmitted and reflected sample paths,
 we can construct the probability distributions for the respective components.
 Figure~\ref{wavemotion} shows that the transmitted wave packet
 is constructed mainly from the front part of the incident wave packet.
 As a result, a displacement of the peak position of the wave packet occurs
 and it looks like its large acceleration.
 This mechanism is similar to the superluminal tunneling of
 the photon \cite{photon1,photon2}.
 We point out the strong dependence of the displacement
on the interaction strength.


\section{Summary}
In this paper, we have investigated the times related to the tunnneling 
of the BEC wave packet in the Nelson's stochastic approach. There are the three
times, namely the tunneling time for which a particle is in a potential barrier,
the hesitating one for which it stays in front of the barrier and the interaction one,
give by a sum of the tunnneling and hesitating times.
 Applying the mean field approach to
 the Nelson's stochastic formulation,
 we derive the GP equation directly.

According to the stochastic formulation, we have performed numerical calculations.
 First, it is found that
 the tunneling time decreases as the interatomic repulsive 
interaction becomes stronger and is not affected so much 
by the potential barrier height
 for the strong interaction.
 The dependence of the hesitating and interaction times on the parameters,
 especially on the interaction strength, has been revealed.
 It is seen that the Hartman effect is violated when the non-linear interaction is switched on.
 
We have also calculated the mean arrival time of the BEC wave packet
 and have seen a large displacement of its peak position.
 This result implies that it is not adequate to define the tunneling
 time by the peak motion ( or the group velocity ) of the BEC wave packet.

\begin{acknowledgments}
One of the author (K.~K.) would like to thank Dr.~M.~Okumura
for a fruitful discussion.
\end{acknowledgments}

\appendix

\newpage 

\end{document}